\documentclass[aps,prd,twocolumn,superscriptaddress,showpacs,preprintnumbers,amsmath,amssymb]{revtex4-1}

\usepackage{graphicx}
\usepackage{dcolumn}
\usepackage{bm}
\usepackage[mathlines]{lineno}
\usepackage{subfigure}
\usepackage{multirow}

\begin{document}

\preprint{\vbox{ \hbox{   }
                  \hbox{Belle-CONF-1602}
}}

\title{Measurement of the branching ratio of $\bar{B}^0 \rightarrow D^{*+} \tau^- \bar{\nu}_{\tau}$
relative to $\bar{B}^0 \rightarrow D^{*+} \ell^- \bar{\nu}_{\ell}$ decays
with a semileptonic tagging method
}



\date{\today}

\noaffiliation
\affiliation{Aligarh Muslim University, Aligarh 202002}
\affiliation{University of the Basque Country UPV/EHU, 48080 Bilbao}
\affiliation{Beihang University, Beijing 100191}
\affiliation{University of Bonn, 53115 Bonn}
\affiliation{Budker Institute of Nuclear Physics SB RAS, Novosibirsk 630090}
\affiliation{Faculty of Mathematics and Physics, Charles University, 121 16 Prague}
\affiliation{Chiba University, Chiba 263-8522}
\affiliation{Chonnam National University, Kwangju 660-701}
\affiliation{University of Cincinnati, Cincinnati, Ohio 45221}
\affiliation{Deutsches Elektronen--Synchrotron, 22607 Hamburg}
\affiliation{University of Florida, Gainesville, Florida 32611}
\affiliation{Department of Physics, Fu Jen Catholic University, Taipei 24205}
\affiliation{Justus-Liebig-Universit\"at Gie\ss{}en, 35392 Gie\ss{}en}
\affiliation{Gifu University, Gifu 501-1193}
\affiliation{II. Physikalisches Institut, Georg-August-Universit\"at G\"ottingen, 37073 G\"ottingen}
\affiliation{SOKENDAI (The Graduate University for Advanced Studies), Hayama 240-0193}
\affiliation{Gyeongsang National University, Chinju 660-701}
\affiliation{Hanyang University, Seoul 133-791}
\affiliation{University of Hawaii, Honolulu, Hawaii 96822}
\affiliation{High Energy Accelerator Research Organization (KEK), Tsukuba 305-0801}
\affiliation{Hiroshima Institute of Technology, Hiroshima 731-5193}
\affiliation{IKERBASQUE, Basque Foundation for Science, 48013 Bilbao}
\affiliation{University of Illinois at Urbana-Champaign, Urbana, Illinois 61801}
\affiliation{Indian Institute of Technology Bhubaneswar, Satya Nagar 751007}
\affiliation{Indian Institute of Technology Guwahati, Assam 781039}
\affiliation{Indian Institute of Technology Madras, Chennai 600036}
\affiliation{Indiana University, Bloomington, Indiana 47408}
\affiliation{Institute of High Energy Physics, Chinese Academy of Sciences, Beijing 100049}
\affiliation{Institute of High Energy Physics, Vienna 1050}
\affiliation{Institute for High Energy Physics, Protvino 142281}
\affiliation{Institute of Mathematical Sciences, Chennai 600113}
\affiliation{INFN - Sezione di Torino, 10125 Torino}
\affiliation{J. Stefan Institute, 1000 Ljubljana}
\affiliation{Kanagawa University, Yokohama 221-8686}
\affiliation{Institut f\"ur Experimentelle Kernphysik, Karlsruher Institut f\"ur Technologie, 76131 Karlsruhe}
\affiliation{Kavli Institute for the Physics and Mathematics of the Universe (WPI), University of Tokyo, Kashiwa 277-8583}
\affiliation{Kennesaw State University, Kennesaw, Georgia 30144}
\affiliation{King Abdulaziz City for Science and Technology, Riyadh 11442}
\affiliation{Department of Physics, Faculty of Science, King Abdulaziz University, Jeddah 21589}
\affiliation{Korea Institute of Science and Technology Information, Daejeon 305-806}
\affiliation{Korea University, Seoul 136-713}
\affiliation{Kyoto University, Kyoto 606-8502}
\affiliation{Kyungpook National University, Daegu 702-701}
\affiliation{\'Ecole Polytechnique F\'ed\'erale de Lausanne (EPFL), Lausanne 1015}
\affiliation{P.N. Lebedev Physical Institute of the Russian Academy of Sciences, Moscow 119991}
\affiliation{Faculty of Mathematics and Physics, University of Ljubljana, 1000 Ljubljana}
\affiliation{Ludwig Maximilians University, 80539 Munich}
\affiliation{Luther College, Decorah, Iowa 52101}
\affiliation{University of Maribor, 2000 Maribor}
\affiliation{Max-Planck-Institut f\"ur Physik, 80805 M\"unchen}
\affiliation{School of Physics, University of Melbourne, Victoria 3010}
\affiliation{Middle East Technical University, 06531 Ankara}
\affiliation{University of Miyazaki, Miyazaki 889-2192}
\affiliation{Moscow Physical Engineering Institute, Moscow 115409}
\affiliation{Moscow Institute of Physics and Technology, Moscow Region 141700}
\affiliation{Graduate School of Science, Nagoya University, Nagoya 464-8602}
\affiliation{Kobayashi-Maskawa Institute, Nagoya University, Nagoya 464-8602}
\affiliation{Nara University of Education, Nara 630-8528}
\affiliation{Nara Women's University, Nara 630-8506}
\affiliation{National Central University, Chung-li 32054}
\affiliation{National United University, Miao Li 36003}
\affiliation{Department of Physics, National Taiwan University, Taipei 10617}
\affiliation{H. Niewodniczanski Institute of Nuclear Physics, Krakow 31-342}
\affiliation{Nippon Dental University, Niigata 951-8580}
\affiliation{Niigata University, Niigata 950-2181}
\affiliation{University of Nova Gorica, 5000 Nova Gorica}
\affiliation{Novosibirsk State University, Novosibirsk 630090}
\affiliation{Osaka City University, Osaka 558-8585}
\affiliation{Osaka University, Osaka 565-0871}
\affiliation{Pacific Northwest National Laboratory, Richland, Washington 99352}
\affiliation{Panjab University, Chandigarh 160014}
\affiliation{Peking University, Beijing 100871}
\affiliation{University of Pittsburgh, Pittsburgh, Pennsylvania 15260}
\affiliation{Punjab Agricultural University, Ludhiana 141004}
\affiliation{Research Center for Electron Photon Science, Tohoku University, Sendai 980-8578}
\affiliation{Research Center for Nuclear Physics, Osaka University, Osaka 567-0047}
\affiliation{RIKEN BNL Research Center, Upton, New York 11973}
\affiliation{Saga University, Saga 840-8502}
\affiliation{University of Science and Technology of China, Hefei 230026}
\affiliation{Seoul National University, Seoul 151-742}
\affiliation{Shinshu University, Nagano 390-8621}
\affiliation{Showa Pharmaceutical University, Tokyo 194-8543}
\affiliation{Soongsil University, Seoul 156-743}
\affiliation{University of South Carolina, Columbia, South Carolina 29208}
\affiliation{Sungkyunkwan University, Suwon 440-746}
\affiliation{School of Physics, University of Sydney, New South Wales 2006}
\affiliation{Department of Physics, Faculty of Science, University of Tabuk, Tabuk 71451}
\affiliation{Tata Institute of Fundamental Research, Mumbai 400005}
\affiliation{Excellence Cluster Universe, Technische Universit\"at M\"unchen, 85748 Garching}
\affiliation{Department of Physics, Technische Universit\"at M\"unchen, 85748 Garching}
\affiliation{Toho University, Funabashi 274-8510}
\affiliation{Tohoku Gakuin University, Tagajo 985-8537}
\affiliation{Department of Physics, Tohoku University, Sendai 980-8578}
\affiliation{Earthquake Research Institute, University of Tokyo, Tokyo 113-0032}
\affiliation{Department of Physics, University of Tokyo, Tokyo 113-0033}
\affiliation{Tokyo Institute of Technology, Tokyo 152-8550}
\affiliation{Tokyo Metropolitan University, Tokyo 192-0397}
\affiliation{Tokyo University of Agriculture and Technology, Tokyo 184-8588}
\affiliation{University of Torino, 10124 Torino}
\affiliation{Toyama National College of Maritime Technology, Toyama 933-0293}
\affiliation{Utkal University, Bhubaneswar 751004}
\affiliation{CNP, Virginia Polytechnic Institute and State University, Blacksburg, Virginia 24061}
\affiliation{Wayne State University, Detroit, Michigan 48202}
\affiliation{Yamagata University, Yamagata 990-8560}
\affiliation{Yonsei University, Seoul 120-749}
  \author{A.~Abdesselam}\affiliation{Department of Physics, Faculty of Science, University of Tabuk, Tabuk 71451} 
  \author{I.~Adachi}\affiliation{High Energy Accelerator Research Organization (KEK), Tsukuba 305-0801}\affiliation{SOKENDAI (The Graduate University for Advanced Studies), Hayama 240-0193} 
  \author{K.~Adamczyk}\affiliation{H. Niewodniczanski Institute of Nuclear Physics, Krakow 31-342} 
  \author{H.~Aihara}\affiliation{Department of Physics, University of Tokyo, Tokyo 113-0033} 
  \author{S.~Al~Said}\affiliation{Department of Physics, Faculty of Science, University of Tabuk, Tabuk 71451}\affiliation{Department of Physics, Faculty of Science, King Abdulaziz University, Jeddah 21589} 
  \author{K.~Arinstein}\affiliation{Budker Institute of Nuclear Physics SB RAS, Novosibirsk 630090}\affiliation{Novosibirsk State University, Novosibirsk 630090} 
  \author{Y.~Arita}\affiliation{Graduate School of Science, Nagoya University, Nagoya 464-8602} 
  \author{D.~M.~Asner}\affiliation{Pacific Northwest National Laboratory, Richland, Washington 99352} 
  \author{T.~Aso}\affiliation{Toyama National College of Maritime Technology, Toyama 933-0293} 
  \author{H.~Atmacan}\affiliation{Middle East Technical University, 06531 Ankara} 
% ...
\collaboration{The Belle Collaboration}

\noaffiliation
\begin{abstract}
We report a measurement of ratio
${\cal R}(D^*) = {\cal B}(\bar{B}^0 \rightarrow D^{*+} \tau^- \bar{\nu}_{\tau})/{\cal B}(\bar{B}^0 \rightarrow D^{*+} \ell^- \bar{\nu}_{\ell})$,
where $\ell$ denotes an electron or a muon.
The results are based on a data sample
containing $772\times10^6$ $B\bar{B}$ pairs
recorded at the $\Upsilon(4S)$ resonance
with the Belle detector at the KEKB $e^+ e^-$ collider.
We select a sample of $B^0 \bar{B}^0$ pairs by reconstructing both $B$ mesons in semileptonic decays to
$D^{*\mp} \ell^{\pm}$.
We measure ${\cal R}(D^*)= 0.302 \pm 0.030({\rm stat)} \pm 0.011({\rm syst)}$,
which is within $1.6 \sigma$ of the Standard Model theoretical expectation,
where $\sigma$ is the standard deviation including systematic uncertainties.
\end{abstract}


\maketitle

\section{INTRODUCTION}
Semitauonic $B$ meson decays of the type $b \rightarrow c \tau \nu_{\tau}$~\cite{CHARGE_CONJUGATION}
are sensitive probes to search for physics beyond the Standard Model (SM).
Charged Higgs bosons, which appear in supersymmetry and other models with at least two Higgs doublets,
may contribute to the decay to due to large mass of the $\tau$ lepton and induce measurable effects in the branching fraction.
Similarly, leptoquarks, which carry both baryon number and lepton number, may also contribute to this process.
The ratio of branching fractions
\begin{eqnarray}
{\cal R}(D^{(*)}) = \frac{{\cal B}(\bar{B} \rightarrow D^{(*)} \tau^- \bar{\nu}_{\tau})}{{\cal B}(\bar{B} \rightarrow D^{(*)} \ell^- \bar{\nu}_{\ell})} \hspace{0.8em}(\ell = e,\mu),
\end{eqnarray}
is typically used instead of the absolute branching fraction of $\bar{B} \rightarrow D^{(*)+} \tau^- \bar{\nu}_{\tau}$, to reduce several systematic uncertainties
such as those on the experimental efficiency, the CKM matrix elements $|V_{cb}|$, and on the form factors.
The SM calculations on these ratios predict
${\cal R}(D^*) = 0.252 \pm 0.003$~\cite{SM_PREDICTION_2} and
${\cal R}(D)   = 0.297 \pm 0.017$~\cite{SM_PREDICTION_1,BELLE_HAD_NEW}
with precision of better than 2\% and 6\% for ${\cal R}(D^*)$ and ${\cal R}(D)$, respectively. 
Exclusive semitauonic $B$ decays were first observed by the Belle Collaboration
\cite{BELLE_INCLUSIVE_OBSERVATION},
with subsequent studies reported by 
Belle \cite{BELLE_INCLUSIVE,BELLE_HAD_NEW},
\mbox{\sl B\hspace{-0.4em} {\small\sl A}\hspace{-0.37em} \sl B\hspace{-0.4em}
{\small\sl A\hspace{-0.02em}R}} \cite{BABAR_HAD_NEW},
and LHCb \cite{LHCB_RESULT} Collaborations.
All results are consistent with each other,
and the average values of Refs.~\cite{BELLE_HAD_NEW,BABAR_HAD_NEW,LHCB_RESULT} have been found to be
${\cal R}(D^*) = 0.322 \pm 0.018 \pm 0.012$
and 
${\cal R}(D) = 0.391 \pm 0.041 \pm 0.028$ \cite{HFAG},
which exceed the SM predictions for ${\cal R}(D^*)$ and ${\cal R}(D)$ by $3.0 \sigma$ and $1.7 \sigma$, respectively.
The combined analysis of ${\cal R}(D^*)$ and ${\cal R}(D)$, taking into account measurement correlations, finds that the deviation is $3.9\sigma$ from the SM prediction.

So far,  measurements of ${\cal R}(D^{(*)})$ at the $B$ factories
have been performed either using a hadronic \cite{BELLE_HAD_NEW,BABAR_HAD_NEW} or an inclusive tagging method \cite{BELLE_INCLUSIVE_OBSERVATION,BELLE_INCLUSIVE}.
Semileptonic tagging methods have been employed for use in studies of $B^- \rightarrow \tau^- \bar{\nu}_{\tau}$ decays, and have been shown to be of similar experimental precision to that of the hadronic tagging method \cite{TAUNU_BELLE_SEMILEP,TAUNU_BABAR_SEMILEP}.
In this paper, we report the first measurement of ${\cal R}(D^*)$ using the semileptonic tagging method.
We reconstruct signal $B^0\bar{B}^0$ events in modes where one $B$ decays semi-tauonically $\bar{B}^0 \rightarrow D^{*+} \tau^- \bar{\nu}_{\tau}$
where $\tau^- \rightarrow \ell^- \bar{\nu}_{\ell} \nu_{\tau}$, 
(referred to hereafter as $B_{\rm sig}$)
and the the other $B$ decays in a semileptonic channel $\bar{B}^0 \rightarrow D^{*+} \ell^- \bar{\nu}_{\ell}$
(referred to hereafter as $B_{\rm tag}$).
To reconstruct normalization $B^0\bar{B}^0$ events, which correspond to the denominator in ${\cal R}(D^*)$,
we use both $B$ mesons decaying to semileptonic decay modes $D^{*\pm} \ell^{\mp} \bar{\nu}_{\ell}\hspace{-1.18em}$ \scalebox{0.25}{$\displaystyle \raise5ex\hbox{(\hspace{1.7em})}$}$\hspace{0.25em}$.

\section{DETECTOR AND MC SIMULATION}
We use the full $\Upsilon(4S)$ data sample
containing $772 \times 10^6$ $B \bar{B}$ pairs
recorded with the Belle detector \cite{BELLE}
at the KEKB $e^+ e^-$ collider \cite{KEKB}.
The Belle detector is a general-purpose magnetic spectrometer
which consists of
a silicon vertex detector (SVD),
a 50-layer central drift chamber (CDC),
an array of aerogel threshold Cherenkov counters (ACC),
time-of-flight scintillation counters (TOF),
and an electromagnetic calorimeter (ECL) comprised of CsI(Tl) crystals.
The devices are located inside a superconducting solenoid coil
that provides a 1.5 T magnetic field.
An iron flux-return located outside the coil
is instrumented to detect $K_L^0$ mesons
and to identify muons (KLM).
The detector is described in detail elsewhere \cite{BELLE}.
 
To determine the acceptance and  probability density functions (PDF)  for
signal $\bar{B}^0 \rightarrow D^{*+} \tau^- \bar{\nu}_{\tau}$,
normalization $\bar{B}^0 \rightarrow D^{*+} \ell^- \bar{\nu}_{\ell}$,
and background processes we use Monte Carlo (MC) simulated events, which are based on the EvtGen event generator \cite{EVTGEN} and the GEANT3 package \cite{GEANT}.
The MC samples for signal $\bar{B}^0 \rightarrow D^{*+} \tau^- \bar{\nu}_{\tau}$ events
are generated using the decay model based on the heavy quark effective theory~\cite{SIG_DECAY_MODEL}.

Background $B \rightarrow D^{**} \ell \nu_{\ell}$ events are simulated with the ISGW \cite{ISGW} model and reweighted to match
the kinematics predicted by the LLSW model \cite{LLSW}.
Here, $D^{**}$ denotes the orbitally excited states, $D_1$, $D_2^*$, $D_1'$, and $D_0^*$. Radially excited states are considered negligible.
The normalization mode $B \rightarrow D^* \ell \nu_{\ell}$ is simulated using HQET, and reweighted
according to the current world average form factor values: $\rho^2 = 1.207 \pm 0.015 \pm 0.021$, $R_1 = 1.403 \pm 0.033$,
and $R_2 = 0.854 \pm 0.020$ \cite{HFAG}.
The sample sizes of the signal, $B\bar{B}$, and continuum $q\bar{q}$ $(q=u,d,s,c)$ production processes
correspond to about 40, 10 and 6 times the integrated luminosity of the on-resonance collision data sample, respectively.

\section{EVENT SELECTION}
Charged particle tracks are reconstructed with the SVD and CDC, and the tracks other than $K_S^0 \rightarrow \pi^+ \pi^-$ daughters are required to originate from near the interaction region.
Electrons are identified by a combination of the specific ionization ($dE/dx$) in the CDC,
the ratio of the cluster energy in the ECL
to the track momentum measured with the SVD and CDC,
the response of the ACC,
the shower shape in the ECL,
and the match between the positions of the shower and the track at the ECL surface.
To recover bremsstrahlung photons from electrons,
we add the four-momentum of each photon detected
within 0.05 radians of the original track direction.
Muons are identified
by the track penetration depth and hit distribution in the KLM.
Charged kaons are identified
by combining information from
the $dE/dx$ in the CDC, 
the flight time measured with the TOF,
and the response of the ACC \cite{PID}.
We do not apply any particle identification criteria on charged pions.

Candidate $K_S^0$ mesons are formed
by combining two oppositely charged tracks
with pion mass hypotheses.
We require the invariant mass to lie within 15 MeV/$c^2$ of the nominal $K^0$ mass \cite{PDG}.
We then impose the following additional requirements:
(1) the two pion tracks must have a large distance of closest approach to the IP
in the plane perpendicular to the electron beam line;
(2) the pion tracks must intersect at a common vertex
that is displaced from the IP;
(3) the $K_S^0$ candidate's momentum vector should originate from the IP. 
Neutral pion candidates are formed from pairs of photons
with further criteria specific to whether the $\pi^0$ is from a $D^{*+}$ decay and $D$ decay.
For the neutral pions from $D$ decays,
we require 
the photon daughter energies to be greater than 50 MeV,
the cosine of the angle between two photons to be greater than 0.0,
and the $\gamma \gamma$ invariant mass to be $-15$ to $+10$ MeV/$c^2$ around the nominal $\pi^0$ mass \cite{PDG}
which corresponds to approximately $\pm 1.8 \sigma$,
where photons are measured as an energy cluster
in the ECL with no associated charged tracks.
A mass-constrained fit is then performed to obtain the $\pi^0$ momentum.
For neutral pions from $D^{*+}$ decays, which have lower energies,
we require one photon to have at least 50 MeV and the other to have at least 20 MeV.
We apply a tighter window on the invariant mass to compensate for the lower photon energy requirement,
within 10 MeV/$c^2$ of the nominal $\pi^0$ mass,
which corresponds to approximately $\pm 1.6 \sigma$.

Neutral $D$ mesons are reconstructed in the following decay modes:
$D^0 \rightarrow K^- \pi^+$,
$K_S^0 \pi^0$,
$K^+ K^-$,
$\pi^+ \pi^-$,
$K_S^0 \pi^+ \pi^-$,
$K^- \pi^+ \pi^0$,
$\pi^+ \pi^- \pi^0$,
$K_S^0 K^+ K^-$,
$K^- \pi^+ \pi^+ \pi^-$, and
$K_S^0 \pi^+ \pi^- \pi^0$.
Charged $D$ mesons are reconstructed in the following modes:
$D^+ \rightarrow K_S^0 \pi^+$,
$K^- \pi^+ \pi^+$,
$K_S^0 \pi^+ \pi^0$,
$K^+ K^- \pi^+$, and
$K_S^0 \pi^+ \pi^+ \pi^-$.
The combined reconstructed branching fractions are
37\% and 22\%
for $D^0$ and $D^+$, respectively.
For $D$ decay modes without a $\pi^0$ in the final state, 
we require the invariant mass of the $D$ candidates
to be within 15 MeV/$c^2$ of the $D^0$ or $D^+$ mass,
which corresponds to a window of approximately $\pm 3\sigma$.
For modes with  a $\pi^0$  in the final state,
we require a wider invariant mass window:
from $-45$ to $+30$ MeV/$c^2$ around the nominal $D^0$ mass for $D^0$ candidates
and 
from $-36$ to $+24$ MeV/$c^2$ around the nominal $D^+$ mass for $D^+$ candidates.
Candidate $D^{*+}$ mesons are formed by combining $D^0$ and $\pi^+$ candidates
or $D^+$ and $\pi^0$ candidates.
To improve the resolution of the $D^* - D$ mass difference, $\Delta M$,
the charged pion track from the $D^{*+}$ is refitted to
the $D^0$ decay vertex. We require $\Delta M$ to be within 2.5 MeV/$c^2$ and 2.0 MeV/$c^2$ around
nominal $D^*$-$D$ mass difference
for $D^{*+} \rightarrow D^0 \pi^+$ and $D^{*+} \rightarrow D^+ \pi^0$ decay modes, respectively.
We apply a tighter window in $D^{*+} \rightarrow D^+ \pi^0$ decay mode
to suppress large background from fake neutral pions.

To tag semileptonic $B$ decays,
we combine $D^{*+}$ meson and lepton candidates
of opposite electric charge
and calculate the cosine of the angle between the momentum of the $B$ meson and the $D^* \ell$ system in the $\Upsilon(4S)$ rest frame,
under the assumption that only one massless particle is not reconstructed:
\begin{eqnarray}
\cos \theta_{B \mathchar`-  D^* \ell} \equiv
\frac
{2E_{\rm beam} E_{D^* \ell} - m_B^2 - M_{D^* \ell}^2}
{2 |\vec{p}_B| \cdot |\vec{p}_{D^* \ell}|},
\label{eq:cos_bdstrl}
\end{eqnarray}
where $E_{\rm beam}$ is the energy of the beam, and
$E_{D^* \ell}$, $\vec{p}_{D^* \ell}$ and $M_{D^* \ell}$
are the energy, momentum, and mass of the $D^* \ell$ system, respectively.
The variable $m_B$ is the nominal $B$ meson mass~\cite{PDG}, and $\vec{p}_B$ is the nominal $B$ meson momentum.
All variables are defined in the $\Upsilon(4S)$ rest frame.
Correctly reconstructed $B$ candidates in the tag and normalization mode $D^* \ell \nu_{\ell}$ are expected to have
a value of $\cos \theta_{B \mathchar`-  D^* \ell}$ between $-1$ and $+1$.
On the other hand, correctly reconstructed $B$ candidates in the signal decay mode $D^* \tau \nu_{\tau}$
or falsely reconstructed $B$ candidates would tend to have values of $\cos \theta_{B \mathchar`-  D^* \ell}$
below the physical region
due to contributions from additional particles
and a large negative correlation with missing mass squared,
$M_{\rm miss}^2 = (2E_{\rm beam} - \sum_i E_i)^2/c^4 - |\sum_i \vec{p}_i|^2/c^2$,
where $(\vec{p}_i, E_i)$ is
four-momentum of the particles in the $\Upsilon(4S)$ rest frame.

In each event we require two tagged $B$ candidates that are opposite in flavor.
Signal events may have the same flavor due to the $B \bar{B}$ mixing,
however we veto such events as they lead to ambiguous $D^* \ell$ pair assignment
and larger combinatorial background.
We require that at most one $B$ meson is reconstructed in a $D^+$ mode,
in order to avoid large background from fake neutral pions when forming $D^*$ candidates.
In each signal event we assign the candidate with the lowest value of $\cos \theta_{B \mathchar`-  D^* \ell}$
(referred to hereafter as $\cos \theta_{B \mathchar`-  D^* \ell}^{\rm sig}$)
as $B_{\rm sig}$.
The probability of falsely assigning the $B_{\rm sig}$ as the $B_{\rm tag}$ for signal events is about 3\%.
After the identification of the $B_{\rm sig}$ and $B_{\rm tag}$ candidates, we apply further background suppression criteria.
On the tag side ($B_{\rm tag}$)
we require $-2.0 < \cos \theta^{\rm tag}_{B \mathchar`-  D^* \ell} < +1.5$ in order to select $B \rightarrow D^* \ell \nu_{\ell}$.
On the signal side we require the $D^*$ momentum in the $\Upsilon(4S)$ rest frame to be less than 2.0 GeV/$c$,
while we require it to be less than 2.5 GeV/$c$ on the tag side, which accounts for differing lepton masses.
Finally, we require the events to contain no extra charged tracks, $K_S^0$ candidates, or $\pi^0$ candidates,
which are reconstructed with the same criteria as those used in the $D$ candidates.
At this stage, the probability of finding multiple candidates is 7\%,
and the average number of candidates is 1.08.
When multiple candidates are found in an event,
we select the most signal-like events based on the quality of vertex-constrained fits for the $D$ mesons.

\section{BACKGROUND SUPPRESSION}
To separate reconstructed signal and normalization events,
we employ a neural network approach based on the ``NeuroBayes'' software package
 \cite{NEUROBAYES}.
The variables used as inputs to the network are
(i) $\cos \theta^{\rm sig}_{B \mathchar`-  D^* \ell}$,
(ii) missing mass squared, $M_{\rm miss}^2$, and
(iii) visible energy $E_{\rm vis} = \sum_i E_i$,
where $E_i$ is
energies of the particles in the $\Upsilon(4S)$ rest frame.
The most powerful observable in separating signal and background is $\cos \theta^{\rm sig}_{B \mathchar`-  D^* \ell}$.
The neural network is trained using MC samples of signal and normalization events.

The most dominant background contribution arises from 
events with falsely reconstructed (fake) $D^{(*)}$ mesons.
We categorize events, in which $D^{(*)}$ candidates are falsely reconstructed in any events,
into fake $D^{(*)}$ events.
The next most dominant contributions arise from two sources
in which $D^*$ mesons from both $B_{\rm sig}$ and $B_{\rm tag}$ are correctly reconstructed.
One source is $B \rightarrow D^{**} \ell \nu_{\ell}$,
where the $D^{**}$ decays to $D^{(*)}$ along with
accompanying particles.
The other source is $B \rightarrow X_c D^*$ events,
where one $D^*$ meson is correctly reconstructed
and the other charmed meson $X_c$ decays via a semileptonic mode.
If the hadrons in the semileptonic $X_c$ decay are not identified, such events can mimic signal.
Similarly, when $X_c$ is $D_s^+$ meson which decays into $\tau^+ \nu_{\tau}$,
such events can also mimic signal.
To separate signal and normalization events from background processes,
we use the extra energy, $E_{\rm ECL}$,
which is defined as the sum of the energies of neutral clusters
detected in the ECL that are not associated with reconstructed particles.
To mitigate photons related to beam background in the energy sum,
we only include clusters with energies greater than 50, 100, and 150 MeV
for the barrel, forward, and backward calorimeter regions, respectively.
Signal and normalization events peak near zero in $E_{\rm ECL}$,
while backgrounds populate a wider range.
We require $E_{\rm ECL}$ to be less than 1.2 GeV.

\section{MC CALIBRATION}
To improve the accuracy of the MC simulation we apply a series of calibration factors determined from control sample measurements.
The lepton identification efficiencies are corrected for electrons and muons, respectively, 
to account for differences between the detector responses in data and MC. 
We reweight events to account for differing $D^{(*)}$ yields between data and MC samples.
The differing yields of truly reconstructed $D^{(*)}$ mesons
between data and MC samples affect ${\cal R}(D^*)$ measurements
through the determination of the backgrounds.
It is difficult to precisely estimate the differing yields of falsely reconstructed $D^{(*)}$ mesons between data and MC samples
by only using sideband region in two-dimensional of the $D$ invariant masses ($M_D$) or $\Delta M$.
Therefore, calibration factors for events with both correctly and falsely reconstructed $D$ mesons
are estimated for each $D$ meson sub-decay mode
using a two-dimensional fit to $M_D$.
Precise calibration can be performed
by using samples with two tagged candidates,
which have good purity and are close to final samples for the ${\cal R}(D^*)$ measurement.
A two-dimensional PDF is constructed
by taking the product of the one-dimensional functions
for $M_D$.
The function in each dimension is constructed by the sum of
the signal component
and the background component as modeled by  first-order Chebychev polynomials.
The signal component is modeled by 
a triple Gaussian for $D^0$ decay modes without a $\pi^0$
or
a Crystal Ball function \cite{CB_FUNCTION} plus a Gaussian for $D^0$ decay modes with a $\pi^0$ and $D^+$ decay modes.
In this calibration, we do not distinguish signal and tag side.
To estimate calibration factors for specific $D$ sub-decay modes,
we fit samples in which
one $D$ meson is reconstructed in a specific mode
while the other $D$ meson is reconstructed in any signal mode.
From the signal and background yield ratios of data to MC samples,
we derive calibration factors of the specific sub-decay mode for events with correctly and falsely reconstructed $D$ mesons.
We can not independently determine calibration factors for all $D$ meson sub-decay modes,
as we use other sub-decay modes
when we calibrate one specific sub-decay mode of a given $D$ meson.
To estimate all the calibration factors correctly,
we first perform the two-dimensional fitting for each sub-decay mode separately, then
repeat the process, weighting samples by the estimated calibration factors,
until all calibration factors  converge.
Similarly, we estimate calibration factors for events with correctly and falsely reconstructed $D^*$ mesons
from a two-dimensional fit to $\Delta M$.
Calibration factors for events with correctly and falsely reconstructed $D^*$ mesons
are separately estimated for $D^0$ and $D^+$ mesons.

\section{MAXIMUM LIKELIHOOD FIT}
We extract the signal and normalization yields
using a two-dimensional extended maximum-likelihood fit in
$\mathit{NN}$ and $E_{\rm ECL}$.
The likelihood function consists of five components:
signal, normalization, fake $D^{(*)}$ events, $B \rightarrow D^{**} \ell \nu_{\ell}$,
and other backgrounds predominantly from $B \rightarrow X_c D^*$.
The PDFs of all components are determined based on MC simulation.
There are significant correlations between $\mathit{NN}$ and $E_{\rm ECL}$
in the background components, but not for the signal. 
We therefore construct the background PDFs using two-dimensional histogram PDFs,
and apply a smoothing procedure to account for limited statistical power \cite{SMOOTHING}.
We construct the signal PDF by taking the product of one-dimensional histograms
in $\mathit{NN}$ and $E_{\rm ECL}$.

Three parameters are floated in the final fit, corresponding to
the yields of the signal, normalization, and $B \rightarrow D^{**} \ell \nu_{\ell}$ components.
The yields of fake $D^{(*)}$ events are fixed
to the values
estimated from sidebands in the $\Delta M$ distributions.
Since the PDF shape of fake $D^{(*)}$ events depends on the composition of
signal, normalization, $B \rightarrow D^{**} \ell \nu_{\ell}$, and other backgrounds,
the relative contributions of these processes to the fake $D^{(*)}$ component are described as a function of the
 the three fitting parameters. 
The yields of other backgrounds
are fixed to the values expected from MC simulation.
The ratio ${\cal R}(D^*)$ is derived from the formula:
\begin{eqnarray}
{\cal R}(D^*) &=&
\frac{1}
{
2{\cal B}(\tau^- \rightarrow \ell^- \bar{\nu}_{\ell} \nu_{\tau})
}
\cdot
\frac{\varepsilon_{\rm norm}}{\varepsilon_{\rm sig}}
\cdot
\frac{N_{\rm sig}}{N_{\rm norm}},
\label{eq:cal_rdstr}
\end{eqnarray}
where
$\varepsilon_{\rm sig (norm)}$ and $N_{\rm sig (norm)}$
are reconstruction efficiency and yields
of signal (normalization) events.
The branching ratios of $\tau^- \rightarrow \ell^- \bar{\nu}_{\ell} \nu_{\tau}$ are based on the current world average values~\cite{PDG}.
The ratio of efficiencies, $\varepsilon_{\rm norm}/\varepsilon_{\rm sig}$, is estimated to be $1.289 \pm 0.015$ from MC simulation.
The difference between reconstruction efficiencies of signal and normalization events arises from their distinct lepton momentum distributions, and the differing event criterion on the $D^*$ momenta on the signal side.

We validate the PDFs used in the fitting procedure by analysing various control samples. For fake $D^{(*)}$ events we study the $\Delta M$ sidebands, where we find good agreement in both $\mathit{NN}$ and $E_{\rm ECL}$.
For $B \rightarrow D^* \ell \nu_{\ell}$ decays, we require one $B$ meson to be reconstructed with the hadronic tagging method,
and the other $B$ meson  reconstructed with the nominal criteria of this analysis.
We find good agreement between data and MC in the
$E_{\rm ECL}$, $M_{\rm miss}^2$, and $E_{\rm vis}$ distributions,
while we find small discrepancies in the $\cos \theta_{B \mathchar`-  D^* \ell}$ distributions
and thus include the differences as a systematic uncertainty.

\section{SYSTEMATIC UNCERTAINTIES}
To estimate the systematic uncertainties on ${\cal R}(D^*)$,
we vary all assumed parameters by one standard deviation and repeat the fit taking the resulting change in ${\cal R}(D^*)$.
The systematic uncertainties are summarized in Table~\ref{tab:systematics}.
The dominant systematic uncertainty arises from
the limited size of the MC samples: to estimate this uncertainty,
we recalculated PDFs for signal, normalization, fake $D^{(*)}$ events, $B \rightarrow D^{**} \ell \nu_{\ell}$, and other backgrounds
by generating toy MC samples from the nominal PDFs according to Poisson statistics
and repeated the fit with the new PDFs.
Small discrepancies between the data and MC are found in the $\cos \theta_{B \mathchar`-  D^* \ell}$ distributions in the hadronic tagged samples.
We correct the $\cos \theta_{B \mathchar`-  D^* \ell}$ distribution in MC samples
according to the observed discrepancy
and repeat the fit.
The estimated uncertainty are referred as ``PDF shape of the normalization in $\cos \theta_{B \mathchar`-  D^* \ell}$''
in Table~\ref{tab:systematics}.
The branching ratios of the $B \rightarrow D^{**} \ell \nu_{\ell}$ decay modes and the decays of the $D^{**}$ mesons are not well known,
and therefore they contribute a large uncertainty
in PDF shape of $B \rightarrow D^{**} \ell \nu_{\ell}$.
The branching ratio of each $B \rightarrow D^{**} \ell \nu_{\ell}$ decay
is varied within their uncertainties.
The uncertainties are assumed
to be
$\pm 6\%$ for $D_1$, 
$\pm 10\%$ for $D_2^*$,
$\pm 83\%$ for $D_1'$,
and
$\pm 100\%$ for $D_0^*$,
respectively,
including limited knowledge of the $D^{**}$ decays.
Furthermore,
we consider the impact of contributions from radially excited $D(2S)$ and $D^*(2S)$,
where we consider  the assuming branching ratios of $B \rightarrow D^{(*)}(2S) \ell \nu_{\ell}$
to be as much as $0.5\%$ each. 
The yields of fake $D^{*}$ events are fixed
to the values estimated from sidebands in the $\Delta M$ distributions.
We vary the fixed yields of fake $D^{(*)}$ events
within the uncertainties.
To take into account possible dependence of PDF shape to $D$ meson sub-decay mode,
we vary the calibration factors for each $D$ meson sub-decay mode
within their uncertainties for events with
falsely reconstructed $D^{(*)}$ events.
The yields of other background processes, predominantly from $B \rightarrow X_c D^*$ events, are fixed
to the values estimated from MC simulation. We consider variations on the yield and shape of the PDF of these background processes, corresponding to their  measured uncertainties.
The uncertainties of each $B \rightarrow X_c D^*$ decays are assumed
to be
$\pm  8\%$ for $B \rightarrow D_s^*  D^{*-}$,
$\pm 14\%$ for $B \rightarrow D_s    D^{*-}$,
$\pm  8\%$ for $B \rightarrow D^{*+} D^{*-}$, and
$\pm 10\%$ for $B \rightarrow D^+    D^{*-}$,
respectively.
Furthermore,
we add an uncertainty of $\pm 4\%$ due to the size of the MC sample.
We include an uncertainty on the branching ratio of $D_s \rightarrow \tau \nu_{\tau}$ decay,
which  may peak near the signal in the $E_{\rm ECL}$ distribution: it is found to be negligible.
The reconstruction efficiency ratio of signal to normalization events
is varied within its uncertainty,
which is limited by the size of MC samples for signal events.  
We include other minor systematic uncertainties from two sources.
One is an uncertainty from the parameters
that are used for the reweighting of the semileptonic $B \rightarrow D^{(*(*))} \ell \nu_{\ell}$ decays
from the ISGW model to the LLSW model.
The other is an uncertainty on the branching ratio of
$\tau^- \rightarrow \ell^- \bar{\nu}_{\ell} \nu_{\tau}$ decay \cite{PDG}.
The total systematic uncertainty is estimated by summing the above uncertainties in quadrature.

\begin{table*}[htbp]
\caption{List of relative systematic uncertainties in percent.}
\begin{center}
\begin{tabular}{c|c|c|c} \hline
& \multicolumn{3}{|c}{${\cal R}(D^*)$ [\%]} \\ \hline
Sources & $\ell^{\rm sig} = e,\mu$ & $\ell^{\rm sig} = e$ & $\ell^{\rm sig} = \mu$ \\ \hline
MC statistics for each PDF shape                                                &  2.2\% & 2.5\%  & 3.9\% \\
PDF shape of the normalization in $\cos \theta_{B \mathchar`-  D^* \ell}$       &  $^{+1.1}_{-0.0}$\% & $^{+2.1}_{-0.0}$\%  & $^{+2.8}_{-0.0}$\% \\
PDF shape of $B \rightarrow D^{**} \ell \nu_{\ell}$                             &  $^{+1.0}_{-1.7}$\% & $^{+0.7}_{-1.3}$\% & $^{+2.2}_{-3.3}$\% \\
PDF shape and yields of fake $D^{(*)}$                                          &  1.4\% & 1.6\%  & 1.6\% \\
PDF shape and yields of $B \rightarrow X_c D^*$                                 &  1.1\% & 1.2\%  & 1.1\% \\
Reconstruction efficiency ratio $\varepsilon_{\rm norm}/\varepsilon_{\rm sig}$  &  1.2\% & 1.5\%  & 1.9\% \\ 
Modeling of semileptonic decay                                                  &  0.2\% & 0.2\%  & 0.3\% \\
${\cal B}(\tau^- \rightarrow \ell^- \bar{\nu}_{\ell} \nu_{\tau})$               &  0.2\% & 0.2\%  & 0.2\% \\ \hline
Total systematic uncertainties                                                  &  $^{+3.4}_{-3.5}$\% & $^{+4.1}_{-3.7}$\% & $^{+5.9}_{-5.8}$\% \\ \hline
\end{tabular}
\label{tab:systematics}
\end{center}
\end{table*}

\section{RESULTS}
\begin{figure*}[htb]
\centering
\subfigure[$\mathit{NN}$ distribution]{
\includegraphics*[width=8.5cm]{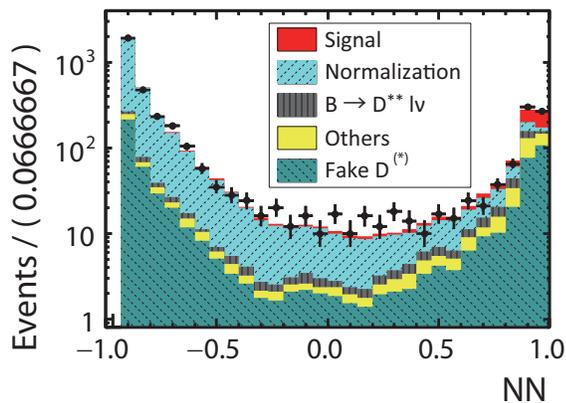}
\label{fig:result_lep0_nb}}
\subfigure[$E_{\rm ECL}$ distribution with signal-enhanced $\mathit{NN}$ region ($\mathit{NN} > 0.8$)]{
\includegraphics*[width=8.5cm]{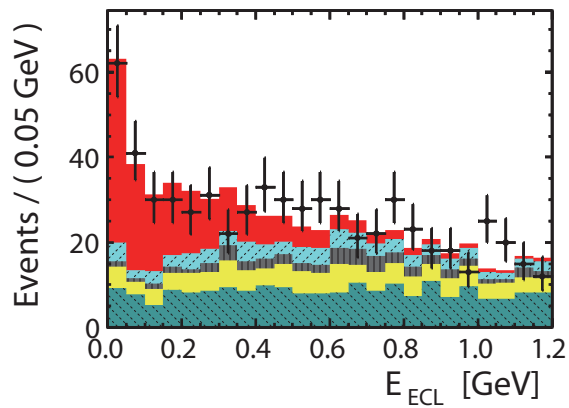}
\label{fig:result_lep0_eecl_sigenh}}
\subfigure[$E_{\rm ECL}$ distribution with normalization-enhanced $\mathit{NN}$ region ($\mathit{NN} < 0.8$)]{
\includegraphics*[width=8.5cm]{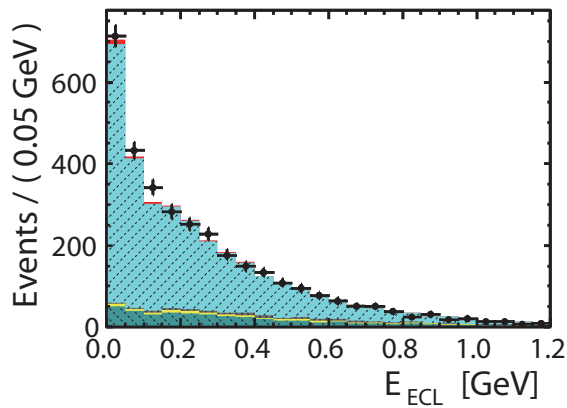}
\label{fig:result_lep0_eecl_normenh}}
\caption{Projections of the fit results with data points overlaid.
The background categories are described in detail in the text,
where ``others'' refers to predominantly $B \rightarrow X_c D^*$ decays.
}
\label{fig:result_fit}
\end{figure*}

The projection of the fitted distributions are shown in Figure~\ref{fig:result_fit}.
The yields of signal and normalization events are measured to be $231 \pm 23({\rm stat})$ and $2800 \pm 57({\rm stat})$, respectively. The ratio
${\cal R}(D^*)$ is therefore found to be
\begin{eqnarray}
{\cal R}(D^*) &=& 0.302 \pm 0.030 \pm 0.011,
\end{eqnarray}
where the first and second errors correspond to statistical and systematic uncertainties, respectively.

We calculate the statistical significance of the signal as $\sqrt{-2\ln ({\cal L}_0/{\cal L}_{\rm max})}$,
where ${\cal L}_{\rm max}$ and ${\cal L}_0$
are the maximum likelihood and the likelihood obtained
assuming zero signal yield, respectively.
We obtain a statistical significance of $13.8\sigma$.
We also estimate the compatibility of the measured value of ${\cal R}(D^*)$
and the SM prediction.
The effect of systematic uncertainties are included
by convolving the likelihood function
with a Gaussian distribution.
We obtain that our result is larger than the SM prediction
by $1.6\sigma$.

\section{CROSS-CHECKS}
To determine the consistency among $\tau$ final states, 
we divide the data samples by lepton flavor on the signal side
and fit them separately.
The efficiency ratios $\varepsilon_{\rm norm}/\varepsilon_{\rm sig}$
are estimated to be
$1.107 \pm 0.016$ and $1.591 \pm 0.030$
for electron and muon channels of the tau decays, respectively.
We obtain
\begin{eqnarray}
{\cal R}(D^*) &=& 0.311 \pm 0.038 \pm 0.013 \hspace{0.5em}(\ell^{\rm sig}=e), \\
{\cal R}(D^*) &=& 0.304 \pm 0.051 \pm 0.018 \hspace{0.5em}(\ell^{\rm sig}=\mu),
\end{eqnarray}
where the first and second errors correspond to statistical and systematic uncertainties, respectively.
The systematic uncertainties are summarized in Table~\ref{tab:systematics}.
These two results are found to be consistent with each other.

To study $B \rightarrow D^{**} \ell \nu_{\ell}$ background contributions,  we require an additional $\pi^0$ in addition to the nominal event selection.
In this control sample,
we calculate $E'_{\rm ECL}$,
which is defined as the remaining energy
after the energy deposit from the additional $\pi^0$ is removed from $E_{\rm ECL}$.
The $B \rightarrow D^{**} \ell \nu_{\ell}$ background contributions are extracted
from them control samples using the nominal fitting method,
replacing  $E_{\rm ECL}$ with $E'_{\rm ECL}$.
We found consistent results for the branching ratios of $B \rightarrow D^{**} \ell \nu_{\ell}$ in the control and signal regions.

\section{NEW PHYSICS COMPATIBILITY TESTS}
We investigated the compatibility of the data samples with type II two-Higgs-doublet model (2HDM) and leptoquark models.
Assuming all neutrinos are left-handed, 
the effective Hamiltonian that contains
all possible four-fermion operators for the $b \rightarrow c \tau \nu_{\tau}$ decay
can be described as follows:
\begin{eqnarray}
{\cal H}_{\rm eff} &=& \frac{4G_F}{\sqrt{2}}V_{cb}
\left[
{\cal O}_{V_1} +
\sum_{X = S_1, S_2, V_1, V_2, T}
C_X {\cal O}_X
\right],
\end{eqnarray}
where the four-Fermi operators, ${\cal O}_X$, are defined as
\begin{eqnarray}
{\cal O}_{S_1} &=& (\bar{c}_L                  b_R) (\bar{\tau}_R                  \nu_{\tau L}), \\
{\cal O}_{S_2} &=& (\bar{c}_R                  b_L) (\bar{\tau}_R                  \nu_{\tau L}), \\
{\cal O}_{V_1} &=& (\bar{c}_L \gamma^{\mu}     b_L) (\bar{\tau}_L \gamma_{\mu}     \nu_{\tau L}), \\
{\cal O}_{V_2} &=& (\bar{c}_R \gamma^{\mu}     b_R) (\bar{\tau}_L \gamma_{\mu}     \nu_{\tau L}), \\
{\cal O}_{T  } &=& (\bar{c}_R \sigma^{\mu \nu} b_L) (\bar{\tau}_R \sigma_{\mu \nu} \nu_{\tau L}),
\end{eqnarray}
and  the $C_X$  parameters correspond to the Wilson coefficients of ${\cal O}_X$.
In the type II 2HDM,
the relevant Wilson coefficient is given as
$C_{S_1} = -m_b m_\tau \tan^2 \beta/m_{H^+}^2$,
where $\tan \beta$ is the ratio of the vacuum expectation values of the two Higgs doublets,
and $m_b$, $m_{\tau}$, and $m_{H^+}$ are the masses of the $b$ quark, $\tau$ lepton, and charged Higgs boson, respectively.
In $\bar{B}^0 \rightarrow D^{*+} \tau^- \bar{\nu}_{\tau}$ decay,
the influence by ${\cal O}_{S_2}$ operator is identical with
that by ${\cal O}_{S_1}$ except for the opposite sign of corresponding Wilson coefficient \cite{SIG_DECAY_MODEL}.
If we consider a contribution from ${\cal O}_{V_1,V2}$ by a new vector boson $W'$,
which couples to left- or right-handed fermion currents,
we must seriously take tight constraints
by the ATLAS~\cite{WPRIME_ATLAS1,WPRIME_ATLAS2} and CMS~\cite{WPRIME_CMS1,WPRIME_CMS2} experiments at the LHC.
Various leptoquark models have been presented to explain anomalies on ${\cal R}(D^{(*)})$ in Ref.~\cite{LQ1}.
Some of leptoquark models generate the tensor operator,
which is the most sensitive operator to $B \rightarrow D^* \tau \nu_{\tau}$ decay.
We choose one representative model, denoted  $R_2$, as a benchmark,
which contains a scalar leptoquark
with quantum numbers $(SU(3)_c, SU(2)_L)_Y = (3,2)_{7/6}$,
where $SU(3)_c, SU(2)_L, Y$ are
the QCD representation,
the weak isospin representation,
and the hypercharge, respectively.
In this leptoquark model,
the relevant Wilson coefficients are related by 
$C_{S_2} = + 7.8 C_T$
at the $b$ quark mass scale,
assuming a leptoquark mass scale of 1 TeV.
$R_2$ type leptoquark model is dedicatedly discussed in Ref.~\cite{LQ2},
because it seems difficult to implement light vector leptoquarks in realistic scenarios
and other types of scalar leptoquark models destabilize proton \cite{PROTON}.

\begin{figure*}[htb]
\centering
\subfigure[Type II 2HDM.]{
\includegraphics*[width=8.5cm]{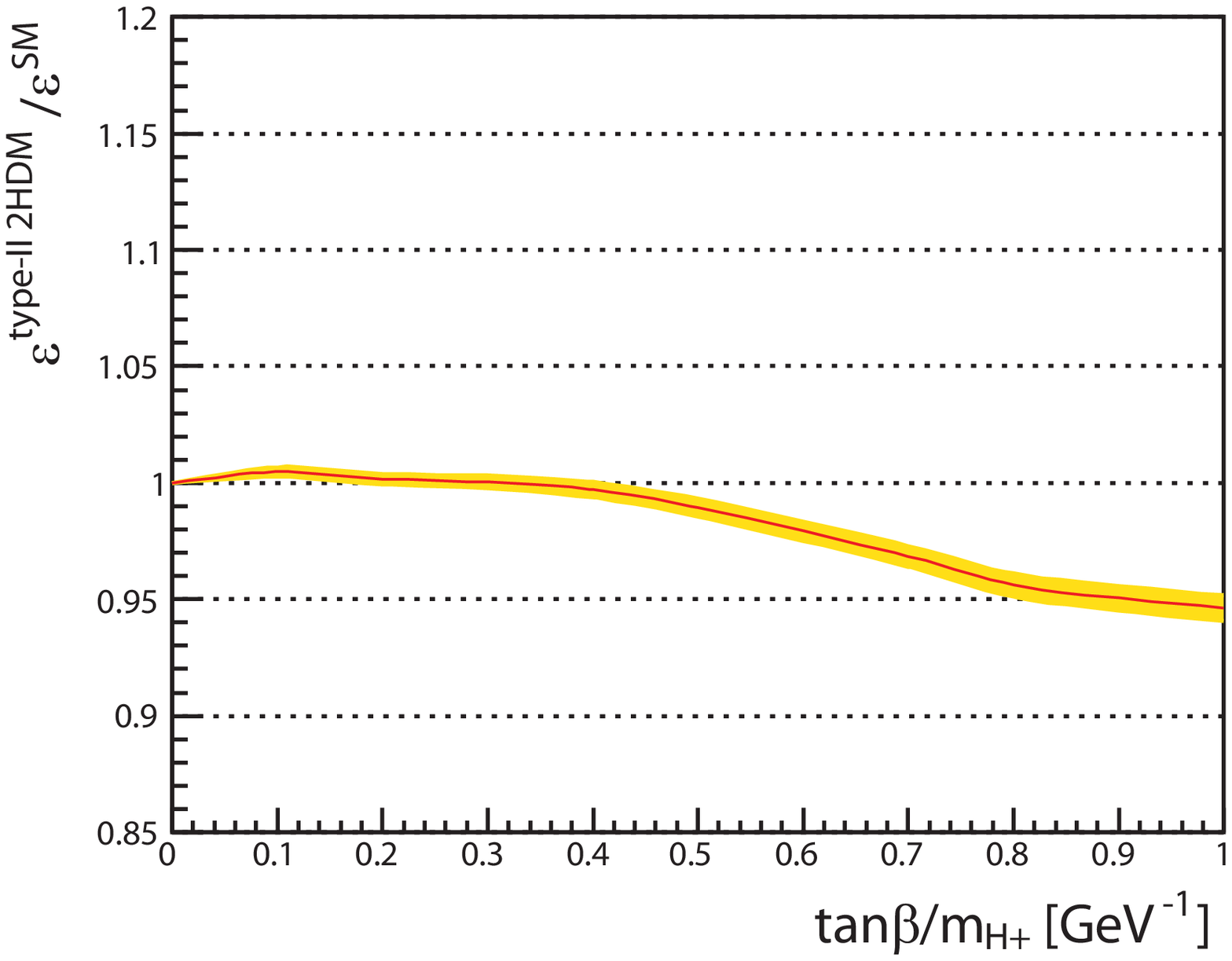}
\label{fig:npcurve_2HDM_eff}}
\subfigure[$R_2$ type leptoquark model.]{
\includegraphics*[width=8.5cm]{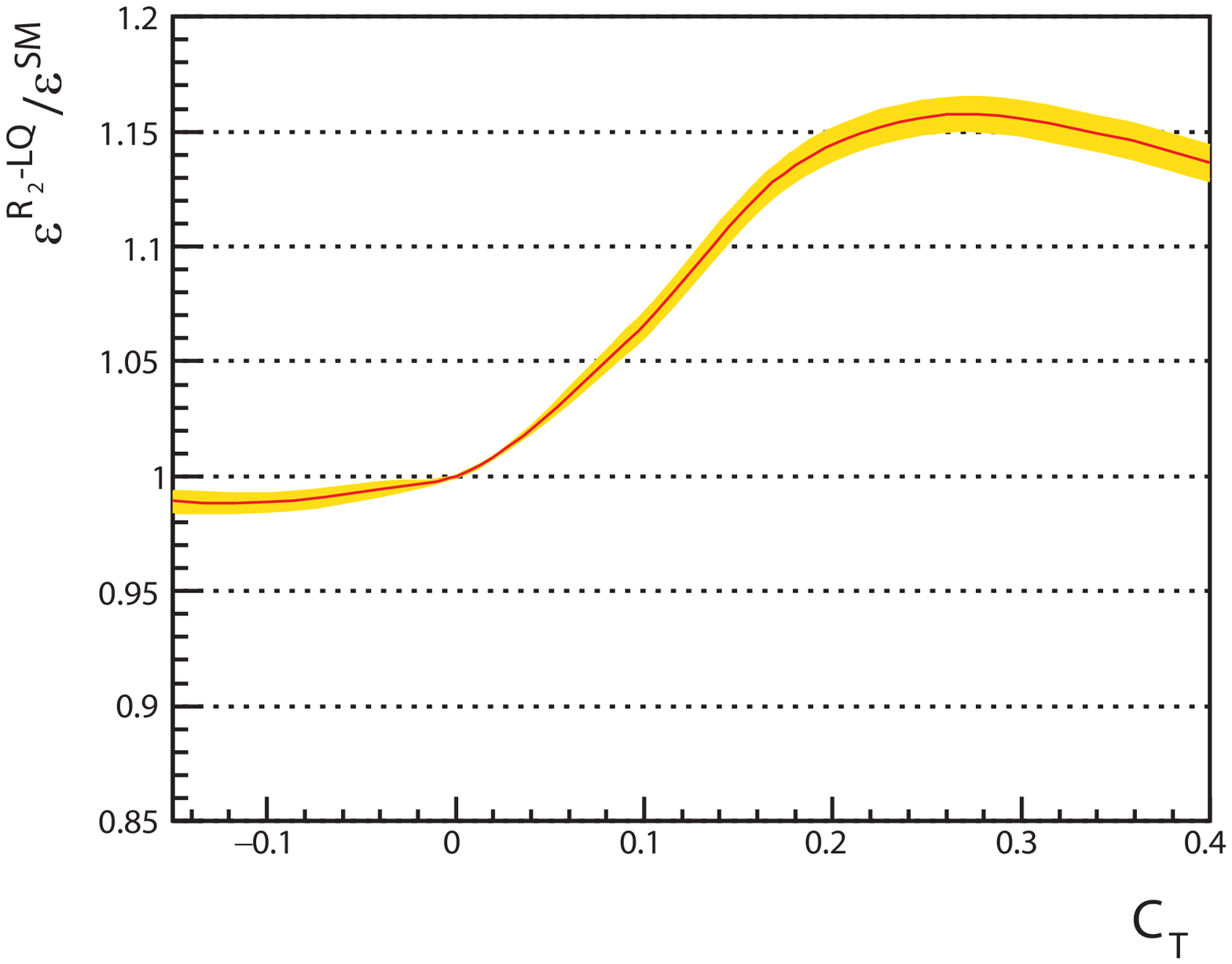}
\label{fig:npcurve_R2LQ_eff}}
\caption{Efficiency in (a) type II 2HDM and (b) $R_2$ type leptoquark model with respect to the SM value.}
\label{fig:npcurve_eff}
\end{figure*}

\begin{figure*}[htb]
\centering
\subfigure[Type II 2HDM.]{
\includegraphics*[width=8.5cm]{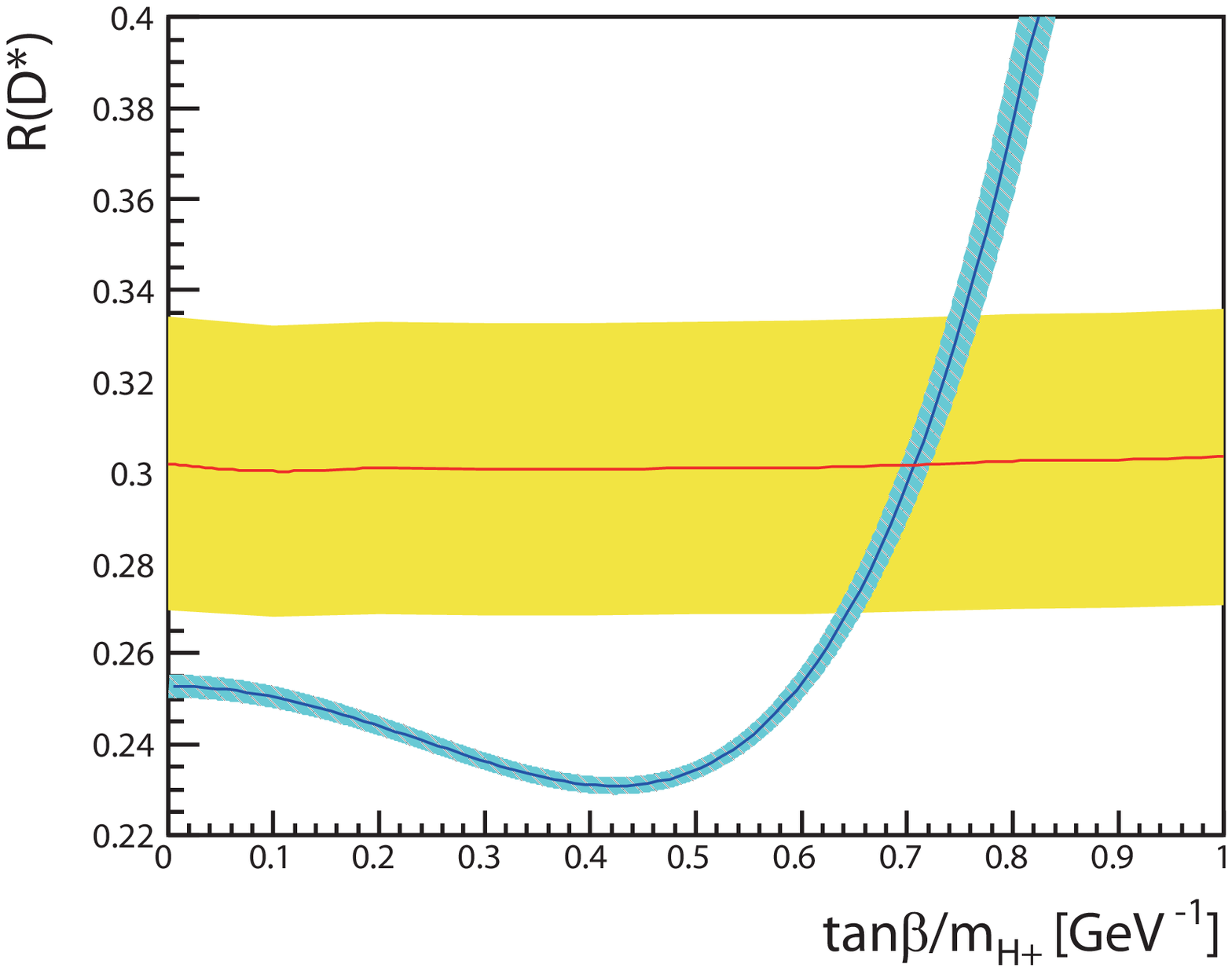}
\label{fig:npcurve_2HDM_rdstr}}
\subfigure[$R_2$ type leptoquark model.]{
\includegraphics*[width=8.5cm]{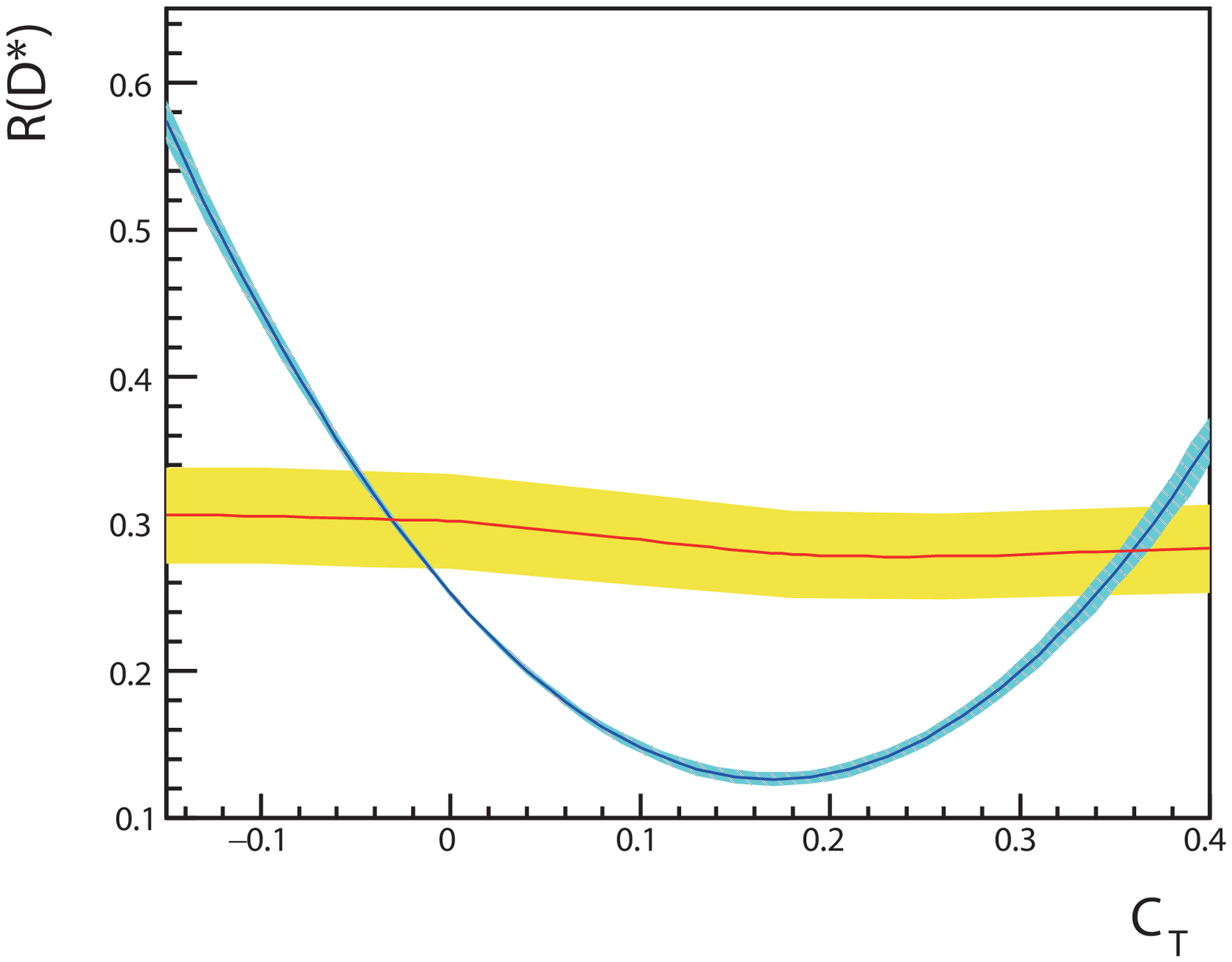}
\label{fig:npcurve_R2LQ_rdstr}}
\caption{
${\cal R}(D^*)$ variation.
Measured values of ${\cal R}(D^*)$ and its uncertainty ($1\sigma$)
in (a) type II 2HDM and (b) $R_2$ type leptoquark model
are shown by solid curve (red) and shaded region (yellow).
Theoretical prediction and its uncertainty ($1\sigma$) are shown by solid curve (blue) and hatched region (light blue)
\cite{SIG_DECAY_MODEL}.}
\label{fig:npcurve_rdstr}
\end{figure*}

To determine the sensitivity to these models, we construct PDFs for signal events
by scanning through values of $\tan\beta/m_{H^+}$ in the type II 2HDM, and $C_T$ in the $R_2$ type leptoquark model.
For the former, $\tan\beta/m_{H^+}$ is scanned
from 0.0 to 1.0 GeV$^{-1}$
and for the latter
$C_T$ is scanned
from $-0.150$ to $+0.400$,
where we assume the Wilson coefficient to be real.
Figure~\ref{fig:npcurve_eff} and \ref{fig:npcurve_rdstr}
demonstrates the dependence of the efficiency and measured values of ${\cal R}(D^*)$
on the values of the respective parameters in the type II 2HDM or the $R_2$ type leptoquark models.
In the type II 2HDM,
the efficiency drops by as much as 5\% for large values of $\tan\beta/m_{H^+}$,
mainly due to the variation of the lepton momentum distribution.
On the other hand,
in the $R_2$ type leptoquark model,
the efficiency increases by up to 16\% at most,
mainly due to the variation of the $D^*$ momentum distribution.
The measured value of ${\cal R}(D^*)$ matches
the theoretical predictions in the type II 2HDM around $\tan \beta / m_{H^+} = 0.7$ GeV$^{-1}$,
while the measured value of ${\cal R}(D^*)$ matches the theoretical predictions
in the $R_2$ type leptoquark model at two points: 
$C_T = -0.03$ and $+0.36$.

\begin{figure*}[htb]
\centering
\subfigure[SM.]{
\includegraphics*[width=5cm]{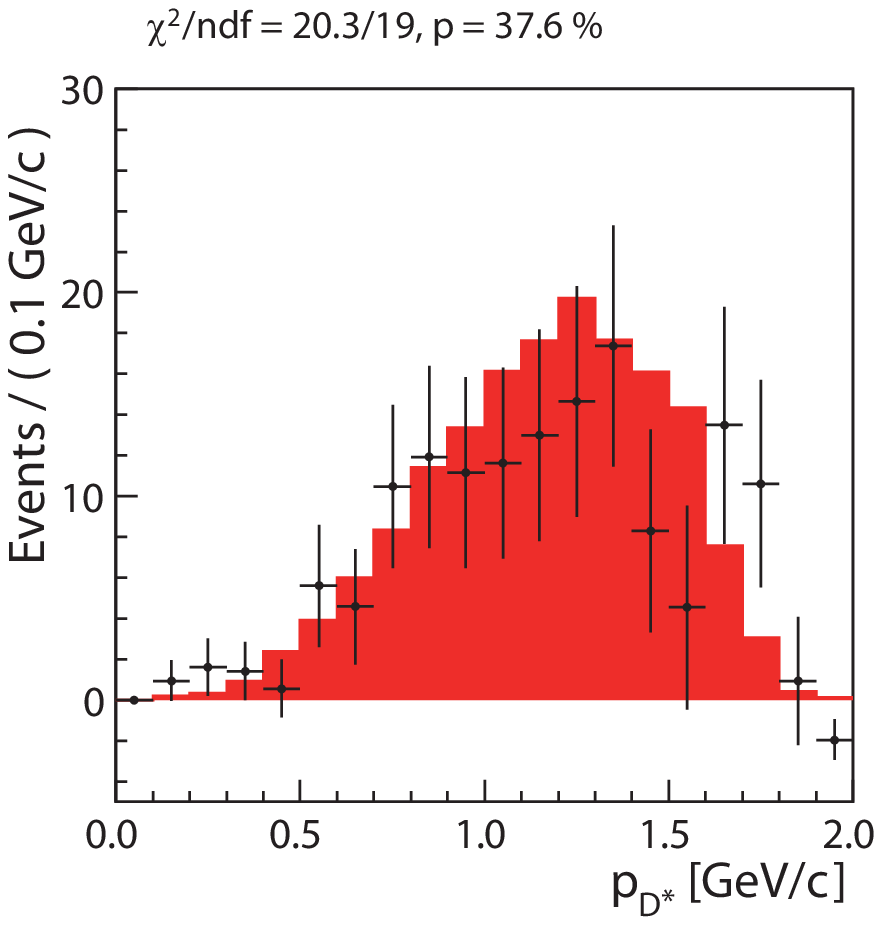}
\label{fig:kinematics_dstr_SM}}
\subfigure[Type II 2HDM with $\tan \beta / m_{H^+} = 0.7$ GeV$^{-1}$.]{
\includegraphics*[width=5cm]{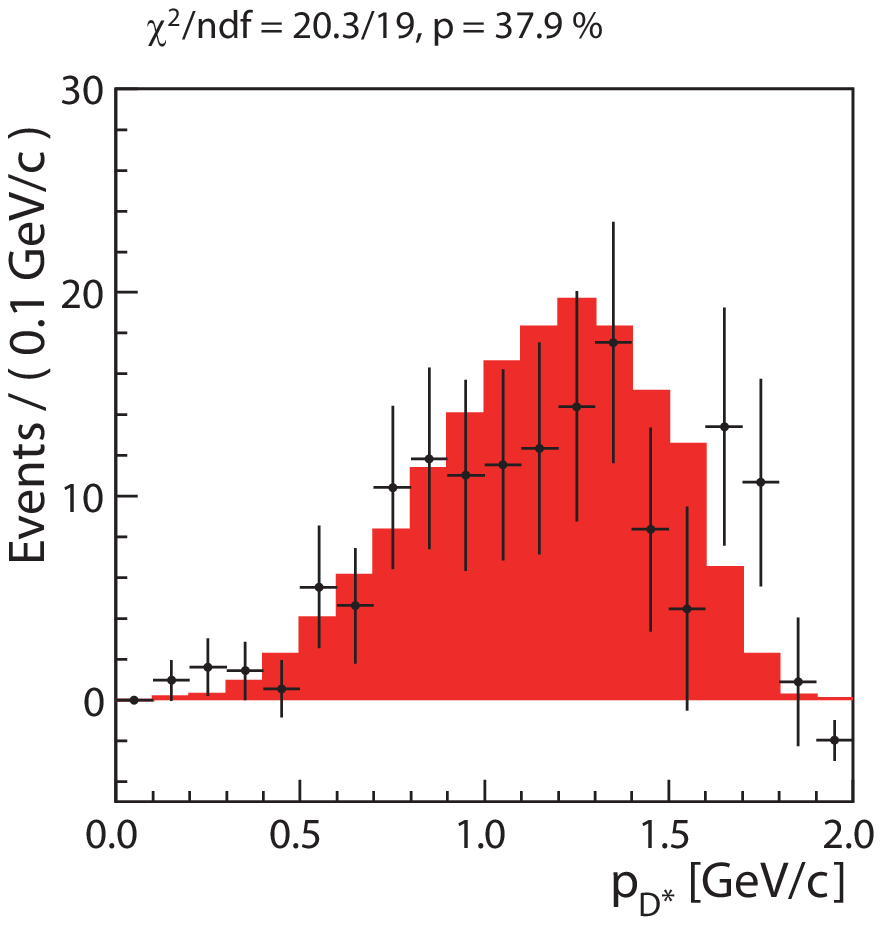}
\label{fig:kinematics_dstr_2hdmII}}
\subfigure[$R_2$ type leptoquark model with $C_T = +0.36$.]{
\includegraphics*[width=5cm]{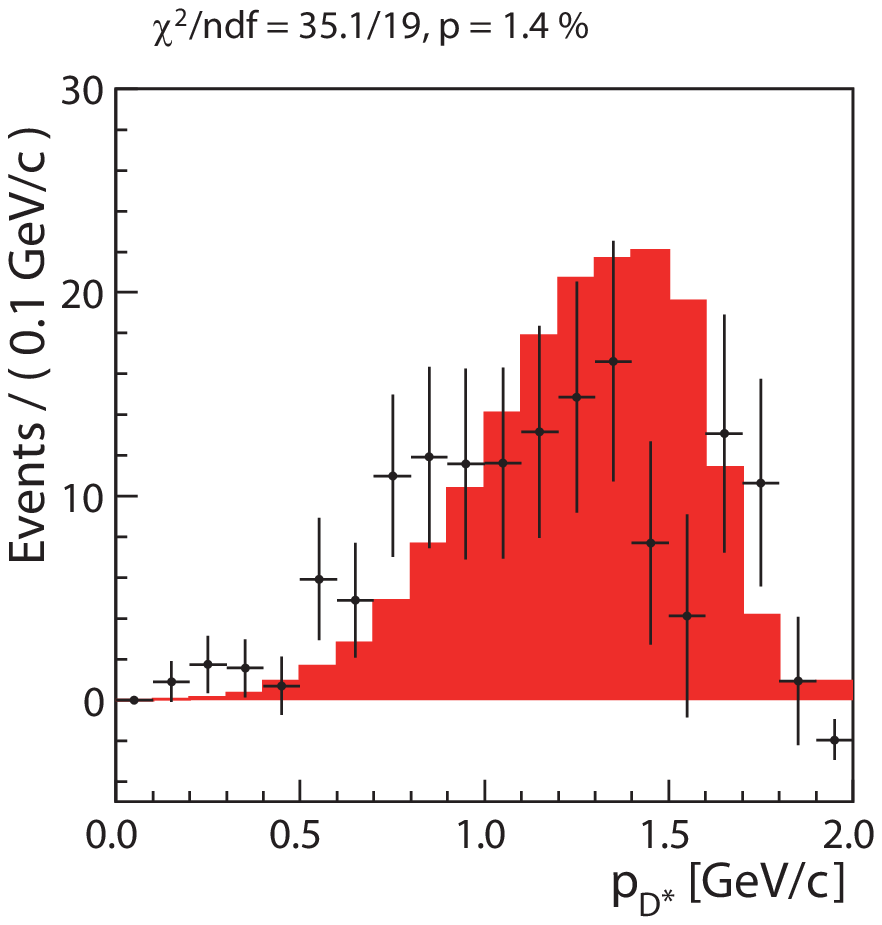}
\label{fig:kinematics_dstr_R2LQ}}
\subfigure[SM.]{
\includegraphics*[width=5cm]{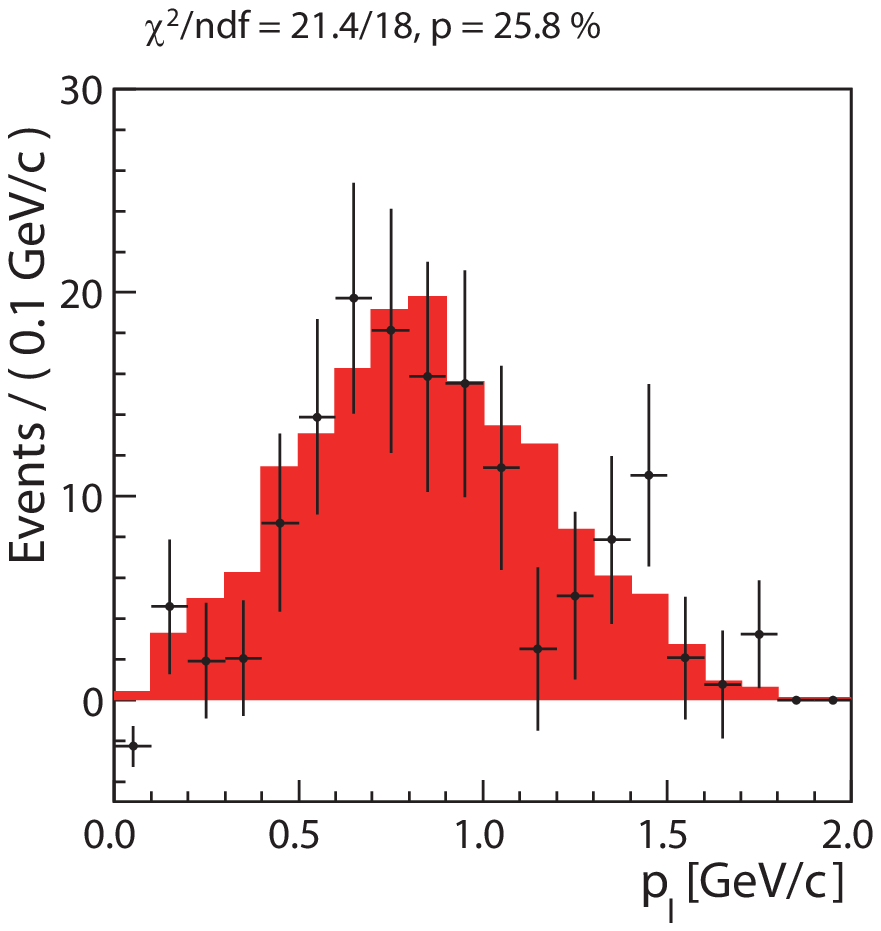}
\label{fig:kinematics_lepton_SM}}
\subfigure[Type II 2HDM with $\tan \beta / m_{H^+} = 0.7$ GeV$^{-1}$.]{
\includegraphics*[width=5cm]{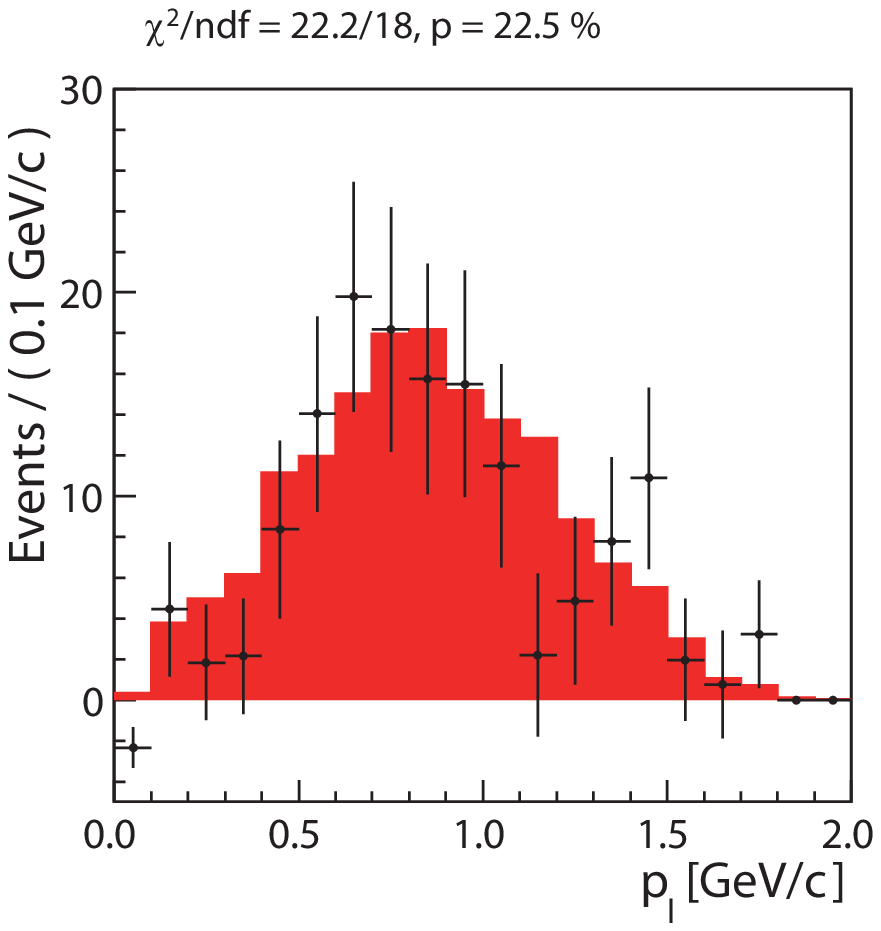}
\label{fig:kinematics_lepton_2hdmII}}
\subfigure[$R_2$ type leptoquark model with $C_T = +0.36$.]{
\includegraphics*[width=5cm]{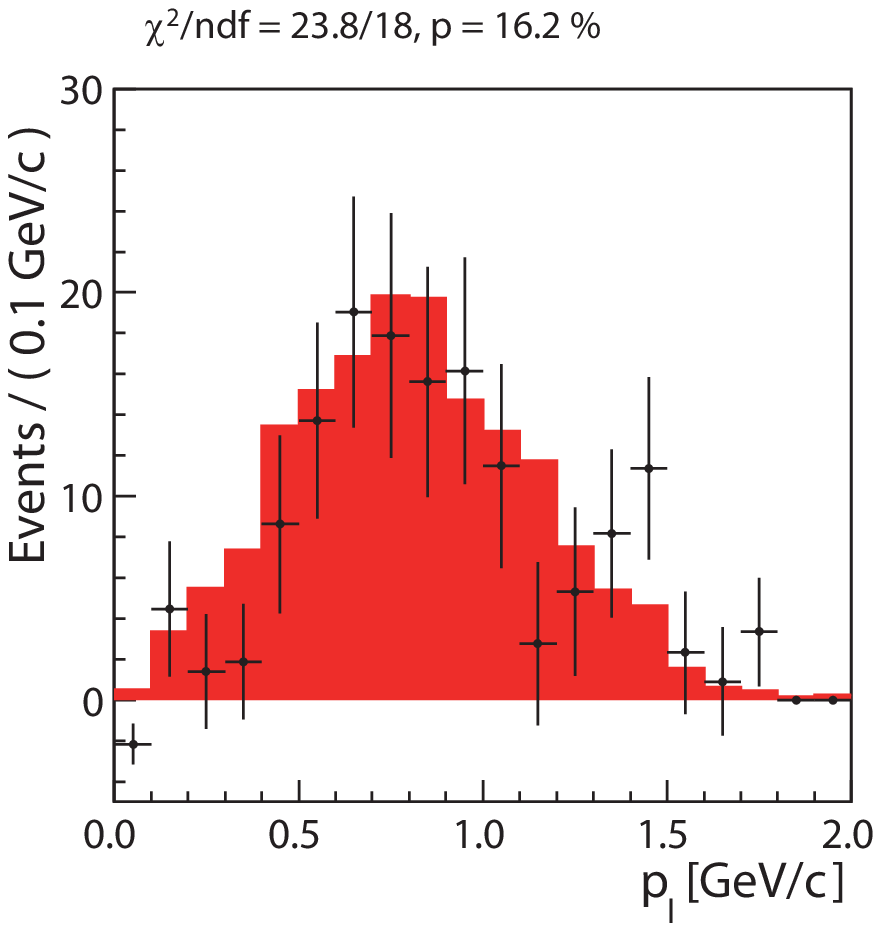}
\label{fig:kinematics_lepton_R2LQ}}
\caption{Background-subtracted momenta distributions of $D^*$ (top) and $\ell$ (bottom)
in the region of $\mathit{NN} > 0.8$ and $E_{\rm ECL} < 0.5$ GeV.
The points and the shaded histograms correspond to the measured and expected distributions, respectively.
The expected distributions are normalized to the number of detected events.
}
\label{fig:kinematics}
\end{figure*}

In Refs.~\cite{BELLE_HAD_NEW} and \cite{BABAR_HAD_NEW},
the $q^2 \equiv (p_B - p_{D^*})^2$ spectra are examined in order to
study the effects of new physics beyond the SM.
Since $q^2$ can not be calculated in this study due to the neutrino from the $B_{\rm tag}$,
we use the momenta of the $D^*$ and the $\ell$ at $B_{\rm sig}$ in $\Upsilon(4S)$ rest frame instead of $q^2$.
Figure~\ref{fig:kinematics} shows the momentum distributions of the background subtracted data
for the SM, type II 2HDM with $\tan \beta / m_{H^+} = 0.7$ GeV$^{-1}$,
and the $R_2$ type leptoquark model with $C_T = +0.36$.
Table~\ref{tab:p_values} shows $p$ values for the three scenarios,
where we include only the statistical uncertainty.
We find our data is compatible with the SM and type II 2HDM with $\tan \beta / m_{H^+} = 0.7$ GeV$^{-1}$,
while the $R_2$ type leptoquark model with $C_T = +0.36$ is disfavored.

\begin{table}[htbp]
\caption{$p$ values for three scenarios.}
\begin{center}
\begin{tabular}{c|c|c|c} \hline \hline
                       &                                         & \multicolumn{2}{c}{$p$ values} \\ \cline{3-4}
Model                  & Parameters                              & $p_{D^*}$ & $p_{\ell}$ \\ \hline
SM                     &                                         & 37.6\%    & 25.8\% \\
Type II 2HDM           & $\tan \beta / m_{H^+} = 0.7$ GeV$^{-1}$ & 37.9\%    & 22.5\% \\
$R_2$ leptoquark model & $C_T = +0.36$                           &  1.4\%    & 16.2\% \\ \hline \hline
\end{tabular}
\label{tab:p_values}
\end{center}
\end{table}

\section{CONCLUSION}
In conclusion,
we report the first measurement of ${\cal R}(D^*)$
with a semileptonic tagging method
using a data sample containing $772 \times 10^6 B\bar{B}$ pairs collected with the Belle detector.
The results are
\begin{eqnarray}
{\cal R}(D^*) &=& 0.302 \pm 0.030({\rm stat}) \pm 0.011({\rm syst}),
\end{eqnarray}
which is within $1.6 \sigma$ of the SM prediction including systematic uncertainties,
is in good agreement with other measurements by Belle~\cite{BELLE_INCLUSIVE_OBSERVATION,BELLE_INCLUSIVE, BELLE_HAD_NEW},
\mbox{\sl B\hspace{-0.4em} {\small\sl A}\hspace{-0.37em} \sl B\hspace{-0.4em}
{\small\sl A\hspace{-0.02em}R}}~\cite{BABAR_HAD_NEW},
and LHCb~\cite{LHCB_RESULT} collaborations,
and is statistically independent of earlier Belle measurements.
We investigate the compatibility
of the data samples with the type II 2HDM and the $R_2$ type leptoquark model.
We find the most favored parameter points are around $\tan \beta / m_{H^+} = 0.7$ GeV$^{-1}$ in the type II 2HDM
and $C_T = -0.030$ and $+0.360$ in the $R_2$ type leptoquark model,
although the latter is disfavored when considering the impact on the decay kinematics.

\section{ACKNOWLEDGEMENTS}
We thank Y.~Sakaki, R.~Watanabe, and M.~Tanaka for their invaluable suggestions.
This work was supported in part by
a Grant-in-Aid for JSPS Fellows (No.13J03438)
and 
a Grant-in-Aid for Scientific Research (S) ``Probing New Physics with Tau-Lepton'' (No.26220706)
We thank the KEKB group for the excellent operation of the
accelerator; the KEK cryogenics group for the efficient
operation of the solenoid; and the KEK computer group,
the National Institute of Informatics, and the 
PNNL/EMSL computing group for valuable computing
and SINET4 network support.  We acknowledge support from
the Ministry of Education, Culture, Sports, Science, and
Technology (MEXT) of Japan, the Japan Society for the 
Promotion of Science (JSPS), and the Tau-Lepton Physics 
Research Center of Nagoya University; 
the Australian Research Council;
Austrian Science Fund under Grant No.~P 22742-N16 and P 26794-N20;
the National Natural Science Foundation of China under Contracts 
No.~10575109, No.~10775142, No.~10875115, No.~11175187, No.~11475187
and No.~11575017;
the Chinese Academy of Science Center for Excellence in Particle Physics; 
the Ministry of Education, Youth and Sports of the Czech
Republic under Contract No.~LG14034;
the Carl Zeiss Foundation, the Deutsche Forschungsgemeinschaft, the
Excellence Cluster Universe, and the VolkswagenStiftung;
the Department of Science and Technology of India; 
the Istituto Nazionale di Fisica Nucleare of Italy; 
the WCU program of the Ministry of Education, National Research Foundation (NRF) 
of Korea Grants No.~2011-0029457,  No.~2012-0008143,  
No.~2012R1A1A2008330, No.~2013R1A1A3007772, No.~2014R1A2A2A01005286, 
No.~2014R1A2A2A01002734, No.~2015R1A2A2A01003280 , No. 2015H1A2A1033649;
the Basic Research Lab program under NRF Grant No.~KRF-2011-0020333,
Center for Korean J-PARC Users, No.~NRF-2013K1A3A7A06056592; 
the Brain Korea 21-Plus program and Radiation Science Research Institute;
the Polish Ministry of Science and Higher Education and 
the National Science Center;
the Ministry of Education and Science of the Russian Federation and
the Russian Foundation for Basic Research;
the Slovenian Research Agency;
Ikerbasque, Basque Foundation for Science and
the Euskal Herriko Unibertsitatea (UPV/EHU) under program UFI 11/55 (Spain);
the Swiss National Science Foundation; 
the Ministry of Education and the Ministry of Science and Technology of Taiwan;
and the U.S.\ Department of Energy and the National Science Foundation.
This work is supported by a Grant-in-Aid from MEXT for 
Science Research in a Priority Area (``New Development of 
Flavor Physics'') and from JSPS for Creative Scientific 
Research (``Evolution of Tau-lepton Physics'').


\begin{thebibliography}{99}

\bibitem{CHARGE_CONJUGATION}
Charge-conjugate decays are implied throughout this paper,
unless otherwise stated.

\bibitem{SM_PREDICTION_2}
S.~Fajfer, J.F.Kamenik, and I.~Nisandzic,
Phys. Rev. D {\bf 85}, 094025 (2012).

\bibitem{SM_PREDICTION_1}
J.F.~Kamenik, and F.~Mescia,
Phys. Rev. D {\bf 78}, 014003 (2008).

\bibitem{BABAR_HAD_NEW}
J.P.~Lees {\it et al.}
(\mbox{\sl B\hspace{-0.4em} {\small\sl A}\hspace{-0.37em} \sl B\hspace{-0.4em}
{\small\sl A\hspace{-0.02em}R}} Collaboration),
Phys. Rev. Lett. {\bf 109}, 101802 (2012);
J.P.~Lees {\it et al.}
(\mbox{\sl B\hspace{-0.4em} {\small\sl A}\hspace{-0.37em} \sl B\hspace{-0.4em}
{\small\sl A\hspace{-0.02em}R}} Collaboration),
Phys. Rev. D {\bf 88}, 072012 (2013);

\bibitem{BELLE_INCLUSIVE_OBSERVATION}
A.~Matyja {\it et al.} (Belle Collaboration),
Phys. Rev. Lett. {\bf 99}, 191807 (2007).

\bibitem{BELLE_INCLUSIVE}
A.~Bozek {\it et al.} (Belle Collaboration),
Phys. Rev. D {\bf 82}, 072005 (2010).


\bibitem{BELLE_HAD_NEW}
M.~Huschle {\it et al.} (Belle Collaboration),
Phys. Rev. D {\bf 92}, 072014 (2015).


\bibitem{LHCB_RESULT}
R.~Aaij {\it et al.} (LHCb Collaboration),
Phys. Rev. Lett. {\bf 115}, 111803 (2015).

\bibitem{HFAG}
Y.~Amhis {\it et al.}, arXiv:1412.7515
and online update at
http://www.slac.stanford.edu/xorg/hfag/

\bibitem{TAUNU_BELLE_SEMILEP}
B.~Kronenbitter {\it et al.} (Belle Collaboration),
Phys. Rev. D {\bf 92}, 051102(R) (2015).


\bibitem{TAUNU_BABAR_SEMILEP}
B.~Aubert {\it et al.}
(\mbox{\sl B\hspace{-0.4em} {\small\sl A}\hspace{-0.37em} \sl B\hspace{-0.4em}
{\small\sl A\hspace{-0.02em}R}} Collaboration),
Phys. Rev. D {\bf 81}, 051101(R) (2010).


\bibitem{BELLE}
A.~Abashian {\it et al.} (Belle Collaboration),
Nucl. Instrum. Methods Phys. Res., Sect. A {\bf 479}, 117 (2002);
also see the detector section in J.~Brodzicka {\it et al.}, Prog. Theor. Exp. Phys. (2012) 04D001.

\bibitem{KEKB}
S.~Kurokawa and E.~Kikutani,
Nucl. Instrum. Methods Phys. Res., Sect. A {\bf 499}, 1 (2003),
and other papers included in this volume;
T.~Abe {\it et al.}, Prog. Theor. Exp. Phys. (2013) 03A001
and following articles up to 03A011.

\bibitem{EVTGEN}
D.J.~Lange Nucl. Instrum. Methods Phys. Res., Sect. A {\bf 462}, 152 (2001).

\bibitem{GEANT}
R.~Brun {\it et al.},
GEAN3.21,
CERN Report No. DD/EE/84-1, 1984 (unpublished)

\bibitem{SIG_DECAY_MODEL}
M.~Tanaka and R.~Watanabe,
Phys. Rev. D {\bf 87}, 034028 (2013).

\bibitem{ISGW}
D.~Scora and N.~Isgur,
Phys. Rev. D {\bf 52}, 2783 (1995).

\bibitem{LLSW}
A.K.~Leibovich, Z.~Ligeti, I.W.~Stewart, and M.B.~Wise,
Phys. Rev. D {\bf 57}, 308 (1998).

 
 
\bibitem{PID}
E.~Nakano,
Nucl. Instrum. Methods Phys. Res., Sect. A {\bf 494}, 402 (2002).

\bibitem{PDG}
Particle Data Group,
Chin. Phys. C {\bf 38}, 090001 (2014).

\bibitem{NEUROBAYES}
M.~Feindt and U.~Kerzel,
Nucl. Instrum. Methods Phys. Res., Sect. A {\bf 559}, 190 (2006).

\bibitem{CB_FUNCTION}
T.~Skwarnicki, Ph.D.~Thesis,
Institute for Nuclear Physics,
Krakow~1986;
DESY Internal Report,
DESY F31-86-02 (1986).


\bibitem{SMOOTHING}
J.H.~Friedman, Data Analysis Techniques for High Energy Particle Physics,
in: Proc. 1974 CERN School of Computing, CERN 74-23 (1974).

\bibitem{WPRIME_ATLAS1}
ATLAS Collaboration, Eur. Phys. J. C {\bf 75},   165 (2015) 

\bibitem{WPRIME_ATLAS2}
ATLAS Collaboration, Phys. Lett. B   {\bf 743},  235 (2015) 

\bibitem{WPRIME_CMS1}
CMS   Collaboration, Phys. Lett. B   {\bf 718}, 1229 (2013) 

\bibitem{WPRIME_CMS2}
CMS   Collaboration, arXiv:1508.04308 

\bibitem{LQ1}
Y.~Sakaki, R.~Watanabe, M.~Tanaka, and A.~Tayduganov,
Phys. Rev. D {\bf 88}, 094012 (2013).

\bibitem{LQ2}
I.~Dor\v{s}ner, S.~Fajfer, N.~Ko\v{s}nik, and I.~Ni\v{s}and\v{z}i\'c,
JHEP, 11, 084 (2013).

\bibitem{PROTON}
I.~Dor\v{s}ner, S.~Fajfer, and N.~Ko\v{s}nik,
Phys. Rev. D {\bf 86}, 015013 (2012).

\end{thebibliography}
\end{document}